\documentclass[sigconf]{acmart}
\usepackage{hyperref}
\usepackage{multirow}
\usepackage{dblfloatfix}
\usepackage{booktabs}
\usepackage{array}
\usepackage{subfigure}
\usepackage[utf8]{inputenc}
\usepackage{algorithm}
\usepackage{algpseudocode}
\usepackage{setspace}
\usepackage{enumitem}
\usepackage{cleveref}
\usepackage{balance} 
\usepackage{mathrsfs}
\usepackage{multirow}
\usepackage{ulem}
\usepackage{makecell} 
\usepackage{graphicx} 
\usepackage{booktabs,multirow,tabularx,threeparttable,xcolor}
\usepackage[table,xcdraw]{xcolor}   

\crefformat{section}{§#2#1#3}

\setlist[itemize]{leftmargin=*, itemsep=0.3em, topsep=0.5em, partopsep=1ex}

\copyrightyear{2026}
\acmYear{2026}
\setcopyright{cc}
\setcctype{by}
\acmConference[KDD '26]{Proceedings of the 32nd ACM SIGKDD Conference on Knowledge Discovery and Data Mining V.2}{August 09--13, 2026}{Jeju Island, Republic of Korea}
\acmBooktitle{Proceedings of the 32nd ACM SIGKDD Conference on Knowledge Discovery and Data Mining V.2 (KDD '26), August 09--13, 2026, Jeju Island, Republic of Korea}
\acmDOI{10.1145/3770855.3818417}
\acmISBN{979-8-4007-2259-2/2026/08}

\AtBeginDocument{%
  }

\begin{document}


\title[From Bootstrapping to Sequence Modeling: A Unified Generative Framework for Personalized \\ Landing-Page Modeling]{From Bootstrapping to Sequence Modeling: A Unified Generative Framework for Personalized Landing-Page Modeling}



\author{Fan Li}
\authornote{Both authors contributed equally to this research.}
\authornote{Work done at Kuaishou Technology.}
\affiliation{%
\institution{Duke University}
\city{Durham}
   \country{USA}
 }
 \email{fan.li@duke.edu}

\author{Chang Meng}
 \authornotemark[1]
\affiliation{%
   \institution{Kuaishou Technology}
   \city{Beijing}
   \country{China}
 }
 \email{mengchang@kuaishou.com}

 \author{Jiaqi Fu}
 \affiliation{%
   \institution{Kuaishou Technology}
   \city{Beijing}
   \country{China}
 }
 \email{fujiaqi05@kuaishou.com}

 \author{Shuchang Liu}
 \affiliation{%
   \institution{Kuaishou Technology}
   \city{Beijing}
   \country{China}
 }
 \email{liushuchang@kuaishou.com}

 \author{Tianke Zhang}
 \affiliation{%
   \institution{Kuaishou Technology}
   \city{Beijing}
   \country{China}
 }
 \email{zhangtianke@kuaishou.com}

 \author{Xueliang Wang}
 \authornote{The corresponding author.}
 \affiliation{%
   \institution{Kuaishou Technology}
   \city{Beijing}
   \country{China}
 }
 \email{wangxueliang03@kuaishou.com}

 \author{Xiaoqiang Feng}
 \authornotemark[3]
 \affiliation{%
   \institution{Kuaishou Technology}
   \city{Beijing}
   \country{China}
 }
 \email{fengxiaoqiang@kuaishou.com}

 \author{Yongqi Liu}
 \affiliation{%
   \institution{Kuaishou Technology}
   \city{Beijing}
   \country{China}
 }
 \email{liuyongqi@kuaishou.com}

 \author{Kaiqiao Zhan}
 \affiliation{%
   \institution{Kuaishou Technology}
   \city{Beijing}
   \country{China}
 }
 \email{zhankaiqiao@kuaishou.com}

\renewcommand{\shortauthors}{Fan Li et al.}

\begin{abstract}
Modern online platforms increasingly adopt multi-page architectures to accommodate diverse user needs.
On these platforms, page navigation (the process of directing users to specific functional pages upon app entry) serves as a critical gateway that shapes user's first impression and significantly influences subsequent engagement. 
To optimize this process, Kuaishou formulated the task of Personalized Landing Page Modeling (PLPM) and proposed KLAN, a reinforcement learning framework built upon Conservative Q-Learning (CQL). 
However, CQL-based approaches suffer from two fundamental limitations: 
(1) the Markov assumption fails to capture the strong non-Markovian temporal dependencies inherent in real-world user behaviors, and 
(2) TD learning with bootstrapping incurs severe cumulative errors and credit assignment difficulties under delayed rewards,
particularly in long-horizon settings where users enter the app multiple times daily.
To address these limitations, we propose GLAN (\underline{\textbf{G}}enerative \underline{\textbf{L}}anding-page \underline{\textbf{A}}daptive \underline{\textbf{N}}avigator), a sequence modeling framework built on Decision Transformer to tackle PLPM from a unified global-local perspective.
Specifically, GLAN incorporates two key modules.
First, we design the L-RTG module that captures users' inter-day consumption dynamics to provide accurate global guidance for all page assignments within a day.
Furthermore, we propose the HRM module that decomposes session-level feedback into fine-grained signals, enabling precise local supervision for each page assignment.
Extensive online experiments conducted on the Kuaishou platform demonstrate the effectiveness of GLAN, achieving \textbf{+0.158\%} and \textbf{+0.108\%} improvements on Daily Active Users (DAU) and user Lifetime (LT) respectively.
\end{abstract}




\maketitle

\begin{figure}[!t]
\includegraphics[width=0.93\linewidth]{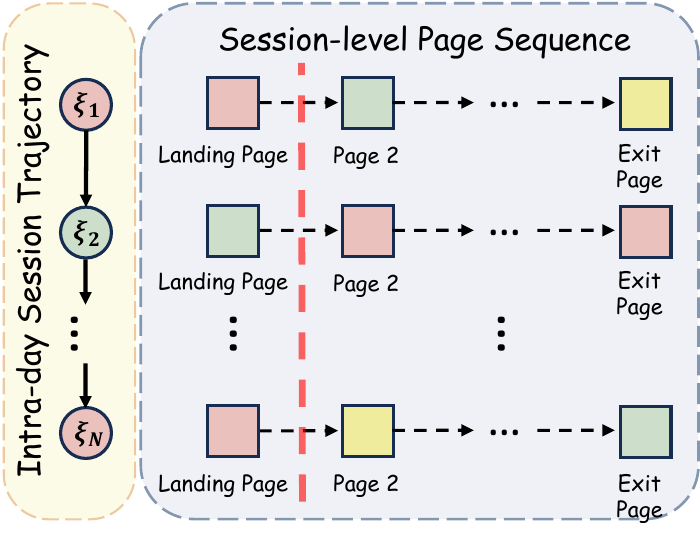}
\vspace{-0.3cm}
\caption{An example of the interaction paradigm in PLPM: 
Within each session $\xi_t$, the landing page is assigned by the platform (left of the red line), while the subsequent pages are actively navigated by the user (right of the red line), forming an ordered page sequence from the landing page to the exit page.
All the  sessions of corresponding user constitute the intra-day trajectory  $\mathcal{T} = (\xi_1, \dots, \xi_N)$.
}
\label{fig:twostage}
\vspace{-0.2cm}
\end{figure}

\begin{figure*}[!t]
    \centering  \includegraphics[width=1.00\linewidth]{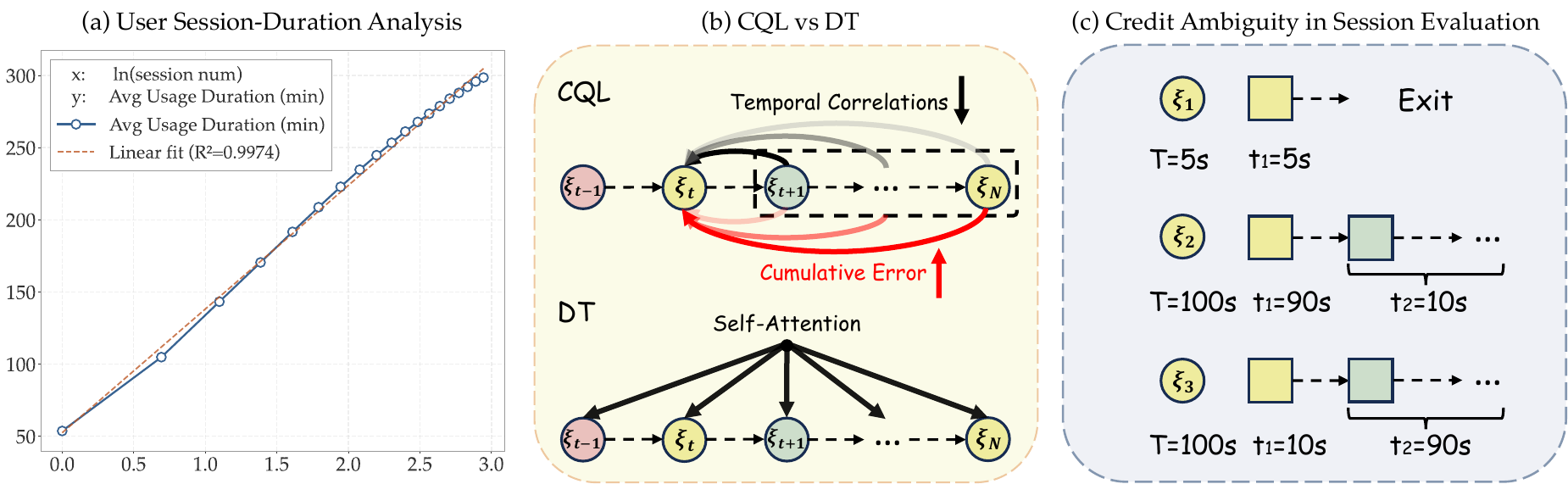}
    \caption{(a) The observed log-linear correlation between daily app usage time and session frequency; (b) the comparison between the step-wise bootstrapping of CQL and the trajectory-based sequence modeling of Decision Transformer; (c) an illustration of the credit ambiguity in naive session-level reward. } 
    \label{fig:DA}
    \vspace{-0.25cm}
\end{figure*}

\section{Introduction}
From e-commerce platforms (e.g., Taobao) to short-video applications (e.g., TikTok and Kuaishou), modern online platforms increasingly adopt a multi-page architecture to cater to users' diverse needs.
In this context, page navigation (the process of navigating users to a specific functional or channel page upon app entry), serves as a critical gateway that shapes the first impression and significantly impacts subsequent user engagement. 
However, with rapidly expanding user bases and increasing industrial constraints, traditional heuristic-based methods struggle to make effective page navigation, often failing to provide fine-grained personalization and adhere to industrial constraints. 
This challenge is further compounded by users' intra-day dynamic interest transition: users typically launch the app multiple times daily with evolving intents, rendering static heuristics fundamentally inadequate.





Recognizing its importance, Kuaishou formulated the task of \textbf{Personalized Landing Page Modeling} (PLPM) ~\cite{li2026klankuaishoulandingpageadaptive} (illustrated in Figure \ref{fig:twostage}) and proposes KLAN~\cite{li2026klankuaishoulandingpageadaptive}, a RL-based framework that models users' intra-day interest dynamics to optimize page assignment upon app entry.
However, the proposed RL-based approach for PLPM suffers from two critical limitations.
First, KLAN adopts CQL~\cite{vanhasselt2015deepreinforcementlearningdouble, kumar2020conservativeqlearningofflinereinforcement, kostrikov2021offlinereinforcementlearningimplicit} based on Markov Decision Processes (MDPs)~\cite{8805177, Zhao_2019, Liu_2023}, yet real-world user behaviors 
exhibit strong non-Markovian temporal dependencies~\cite{10.1145/3726302.3729987, 10.1145/3511808.3557338, chen2021decisiontransformerreinforcementlearning, 10.1145/3543507.3583312, 10.1145/3580305.3599838, 10.1145/3606369, 10.1145/3583780.3615004}. 
Although prior works~\cite{antaris2021sequenceadaptationreinforcementlearning, kang2018selfattentivesequentialrecommendation, liu2019deepreinforcementlearningbased, 10.1145/3690624.3709278, 10.1145/3626772.3657930} attempt to encode historical sequences into the state representation (e.g., using RNNs, Transformers, or mean pooling), such approaches essentially perform lossy compression of trajectory history. Constructing a state representation that strictly satisfies the Markov property remains theoretically challenging and practically intractable in complex industrial environments.
Second, PLPM inherently involves delayed rewards~\cite{weerakoon2022htronefficientoutdoornavigationsparse, ghasemi2025comprehensivesurveyreinforcementlearning, han2021offpolicyreinforcementlearningdelayed}. For instance, assigning a high-stickiness page (e.g., the Following page) in the morning may not yield immediate engagement gains, but can significantly increase the probability of evening revisits and boost the corresponding consumption metrics.
Using temporal difference (TD) ~\cite{10.1023/A:1022633531479} updates to learn a value function through bootstrapping, CQL is prone to severe cumulative errors and credit assignment difficulties under such delayed reward structures. 
These issues are further amplified in long-horizon settings where users enter the app multiple times within a day.

%


%

Recently, Decision Transformer~\cite{10.1145/3726302.3729987, chen2021decisiontransformerreinforcementlearning} has emerged as a promising paradigm, effectively capturing temporal dependencies and historical context via condition sequence modeling.
Applying DT to PLPM offers a principled solution to the aforementioned challenges.
Specifically, to address the non-Markovian issue, DT inherently operates on the full trajectory sequence $(R_1,s_1,a_1,\dots,R_t,s_t,a_t)$, leveraging the Transformer's self-attention mechanism to capture long-range temporal dependencies without requiring the Markov assumption (shown in Figure \ref{fig:DA}(b)).
Moreover, DT circumvents the instability of TD-learning by adopting a supervised learning paradigm conditioned on Returns-to-Go (RTG), effectively avoiding the cumulative error propagation inherent in bootstrapping-based value estimation.
With the global modeling capacity of self-attention and the trajectory-level RTG supervision, DT demonstrates robustness to reward redistribution and outperforms TD-based methods in scenarios requiring long-horizon credit assignment~\cite{brandfonbrener2023doesreturnconditionedsupervisedlearning, bhargava2024preferdecisiontransformersoffline}.






However, several critical challenges emerge when implementing a DT-based approach for PLPM in real-world scenarios.
\begin{itemize}[topsep=4pt]
    \item \textbf{Inter-day RTG Modeling:} 
    During offline training, DT is supervised using posterior RTG values collected from Intra-day trajectory. 
    At inference time, however, the user's future engagement is unavailable, necessitating a prior target RTG to guide global decision-making.
    This introduces a critical initialization dilemma inherent to the DT paradigm:
    conditioning on an overly aggressive target for low-activity users leads to out-of-distribution (OOD) states and model hallucination, whereas overly conservative targets fail to exploit the full potential of high-value users.
    This is further compounded by the inherent periodicity and volatility of user engagement patterns (e.g., weekday vs. weekend behaviors), as well as evolving platform strategies. 
    \item \textbf{Session-level Value Modeling:}
    A naive session-level reward (e.g., total session duration) is not a sufficient statistic for evaluating the landing-page action, resulting in severe credit ambiguity (Figure \ref{fig:DA}(c)). 
    Consider two sessions with identical total duration: 
    in one, the user engages deeply with the assigned landing page, indicating accurate preference alignment; 
    in the other, the user quickly drops the landing page but compensates through downstream consumption on subsequent pages.
    Despite opposite action quality, both yield the same posterior session duration.
    This ambiguity necessitates decomposing session-level feedback into fine-grained components, enabling precise local supervision for each page assignment.
\end{itemize}

To address the aforementioned challenges, we propose GLAN (\underline{\textbf{G}}enerative \underline{\textbf{L}}anding-page \underline{\textbf{A}}daptive \underline{\textbf{N}}avigator), an enhanced DT framework designed to provide personalized landing pages under the formulation of PLPM. 
First, to address the challenge of Inter-day RTG Modeling, we introduce the L-RTG (Lagrangian-constrained Return-to-Go) module.
Recognizing that user engagement exhibits both periodicity and variability, we decouple the modeling into two complementary perspectives: (1) \textit{periodicity-aware patterns}, which capture stable weekly cyclic behaviors, and (2) \textit{sequential dynamics}, which model the evolving and stochastic fluctuations in user interests.
Furthermore, leveraging the inherent synergy between app usage time and session frequency in PLPM (Figure~\ref{fig:DA}(a)), we formulate L-RTG as a constrained optimization problem.
Second, to mitigate the credit ambiguity in Session-level Value Modeling, we design the HRM (Hierarchical Reward Model) module.
Unlike naive duration-based metrics, our model decomposes session-level feedback into fine-grained signals, including page-specific consumption duration and landing-page drop-off risk, enabling the model to distinguish genuine preference alignment from compensatory downstream consumption and thereby providing precise local supervision for each page assignment.

Our contributions can be summarized as follows:
\begin{itemize}[topsep=4pt]
    \item We provide a systematic analysis of the limitations in existing PLPM approaches, examining the compatibility between sequence modeling and traditional RL frameworks.
    \item We propose GLAN, a sequence modeling framework based on Decision Transformer to tackle PLPM from a unified global-local perspective.
    \item We conducted extensive online experiments on the Kuaishou platform, obtaining \textbf{0.158\%} and \textbf{0.108\%} improvements on DAU and LT respectively.
\end{itemize}

\vspace*{-0.5cm}
\section{Related Work}
\subsection{Offline Reinforcement Learning and Decision Transformers}
Unlike traditional methods that optimize for one-step recommendations, reinforcement learning~\cite{sutton1998reinforcement} formulates video recommendation as a Markov Decision Process (MDP), aiming to maximize long-term user satisfaction through cumulative reward optimization.
Reinforcement learning (RL) approaches can be broadly categorized into on-policy~\cite{rummery1994on,mnih2016asynchronous} and off-policy~\cite{haarnoja2018soft,watkins1989learning,vanhasselt2015deepreinforcementlearningdouble} methods.
Due to the high variance and low sample efficiency of on-policy methods, off-policy methods have become the predominant choice in industrial applications.
Within the off-policy paradigm, policy gradient methods~\cite{chen2019large, chen2019topk, ge2021towards, ge2022toward, li2022autolossgen, xian2019reinforcement} suffer from distribution mismatch between the behavior policy and the target policy, leading to substantial discrepancies between offline evaluation and online performance. 
Value gradient approaches~\cite{vanhasselt2015deepreinforcementlearningdouble,pei2019value, taghipour2007usage, zhao2021dear, zhao2018recommendations}, prominently represented by Q-learning~\cite{vanhasselt2015deepreinforcementlearningdouble, kumar2020conservativeqlearningofflinereinforcement, kostrikov2021offlinereinforcementlearningimplicit}, have demonstrated superior stability and data efficiency by leveraging experience replay.
However, these methods share a fundamental limitation: their reliance on the Markov assumption restricts access to historical observations, thereby hindering the modeling of long-horizon dependencies~\cite{bhargava2024preferdecisiontransformersoffline, kumar2022preferofflinereinforcementlearning, brandfonbrener2023doesreturnconditionedsupervisedlearning}, critical in sequential tasks with strong temporal patterns.
Decision Transformers (DT) overcome these limitations by reformulating reinforcement learning as a sequence modeling problem. By leveraging transformer architectures, DT effectively captures historical patterns and long-term dependencies, achieving state-of-the-art performance in offline RL settings. 
Subsequent extensions have further enhanced DT's capabilities, including 
architectural improvement~\cite{liu2023constraineddecisiontransformeroffline, yamagata2023qlearningdecisiontransformerleveraging, Zheng_2025_DecisionMixer, Lu2025VLM}
 and industrial adaptation~\cite{10.1145/3726302.3729987, li2025generativeautobiddinglargescalecompetitive}, demonstrating the versatility of the sequence modeling paradigm for decision-making problems.

\subsection{Page Navigation}
Modern online platforms increasingly adopt a multi-page architecture~\cite{10.1145/3711896.3737275, wu2025adaptivegradientmasking, yu2025whoyouarematters, 10.1145/3637528.3671514, cui2025calibrating, li2025contrastive, 10.1145/3705328.3748070, 10.1145/3770854.3783933} to accommodate diverse user needs, establishing a two-stage interaction paradigm: an initial page navigation stage and a subsequent in-page interaction stage.
Personalized Landing Page Modeling (PLPM)~\cite{li2026klankuaishoulandingpageadaptive} is formulated as a decision-making task that addresses the first stage, 
aiming to proactively select the optimal landing page (e.g., functional tabs or content channels) that maximizes both short-term satisfaction and long-term engagement under industrial constraints.
Early efforts to address PLPM relied on static heuristics, such as resuming the last exited page or selecting the most frequently visited page from offline statistics. 
These approaches lack real-time adaptability and fine-grained personalization, often failing to capture users' evolving intents. 
To bridge this gap, KLAN~\cite{li2026klankuaishoulandingpageadaptive} was proposed as the first framework specifically designed for PLPM, built upon Conservative Q-Learning (CQL)~\cite{kumar2020conservativeqlearningofflinereinforcement}. Additionally, KLAN also introduce Page Drop-off Ratio (PDR)~\cite{li2026klankuaishoulandingpageadaptive} as a key metric to quantify the probability of immediate user departure, serving as a direct proxy for landing page relevance and user satisfaction.

\vspace*{-3pt}
\section{Preliminary}
\subsection{Personalized Landing Page Modeling}
To comprehensively capture the evolution of user interests within a day,
we characterize the interaction paradigm in PLPM as a hierarchical structure 
(as illustrated in Figure~\ref{fig:twostage}).
\begin{itemize}[topsep=4pt]
    \item \textbf{Session-level Page Sequence:}
    A session is defined as the sequence of interactions that starts with the display of a landing page when the user enters the app and ends when the user exits.
    Formally, let $\xi_t$ denote the $t$-th session of a user within a day.
    Each session corresponds to an ordered sequence of visited pages:
    \begin{equation}
    \setlength{\abovedisplayskip}{0.5pt}
\setlength{\belowdisplayskip}{0.5pt}
    \xi_t = (p_{t,1} \rightarrow p_{t,2} \rightarrow \dots \rightarrow p_{t,k} \rightarrow \dots \rightarrow p_{t,M_t}),
    \end{equation}
    where $p_{t,1}$ denotes the \textbf{landing page} and the subsequent pages $\{p_{t,k}\}_{k>1}$ represent the user's voluntary navigation behaviors following the landing page. 
    Here, $M_t$ denotes the total number of pages visited within the $t$-th session.

    \item \textbf{Intra-day Trajectory:}
    Users typically engage with an app across multiple sessions throughout a day, with their intent and interest evolving dynamically over time. 
    Rather than being independent, these sessions exhibit significant temporal dependencies.
    To capture such cross-session temporal dependencies, we define the intra-day trajectory $\mathcal{T}$ as an ordered sequence of sessions. 
    \begin{equation}
    \setlength{\abovedisplayskip}{0.5pt}
\setlength{\belowdisplayskip}{0.5pt}
    \mathcal{T} = (\xi_1, \dots, \xi_t, \xi_{t+1},\dots, \xi_N),
    \end{equation}
    where $N$ denotes the total number of sessions initiated by the user within a day.
\end{itemize}


\subsection{DT-based PLPM}
Decision Transformer (DT)~\cite{chen2021decisiontransformerreinforcementlearning,10.1145/3726302.3729987} has recently emerged as a powerful paradigm for sequential decision-making by reformulating reinforcement learning as a sequence modeling problem. 
Without relying on Markovian assumption~\cite{10.1145/3726302.3729987, 10.1145/3511808.3557338, chen2021decisiontransformerreinforcementlearning}, DT excels at capturing long-range dependencies from a global perspective~\cite{bhargava2024preferdecisiontransformersoffline}, making it well-suited for PLPM where user behaviors display significant temporal dependencies across sessions.
Building upon this insight, we formulate PLPM as a sequential modeling task under DT settings.
\begin{itemize}[topsep=4pt]
    \item \textbf{state} $s_t$: The continuous representation space of the user state. Each
state $s_t$ encodes the user’s real-time features, daily features
upon app entry at step $t$. 
    \item \textbf{action} $a_t$: Each action $a_t$ represents the assignment of a specific
landing page upon app entry at step $t$. The action space $A$ comprises all available landing page (Following Page, Explore Page and Featured Page) assignments. 
    \item \textbf{reward} $\hat{r_t}$: The reward $\hat{r_t}$ is defined as the Session-level value estimate of corresponding session.
    \item \textbf{Return-To-Go (RTG)} $\hat{R_t}$: The RTG value indicates the total
amount of rewards to be obtained in the future time steps:
\begin{equation}
\setlength{\abovedisplayskip}{0.5pt}
\setlength{\belowdisplayskip}{0.5pt}
    \hat{R}_t = \sum \hat{r}_t
\end{equation}

These settings result in the following trajectory representation,
which is well-suited for autoregressive training and inference:
\begin{equation}
\setlength{\abovedisplayskip}{0.5pt}
\setlength{\belowdisplayskip}{0.5pt}
    \mathbf{T} = (\hat{R}_1,s_1,a_1,\hat{R}_2,s_2,a_2\dots,\hat{R}_t,s_t,a_t)
    \vspace{-0.1cm}
\end{equation}
\end{itemize}

\vspace*{-0.2cm}
\subsection{The L-RTG Modeling}
The collected dataset is denoted as $\mathcal{Z} = \{z_1, z_2, \dots, z_N\}$ and $N$ is the total number of instances.
Each instance $z_i = (\mathbf{x}_i, \mathbf{t}_i, \mathbf{se}_i, y_i^{wt}, y_i^{session}) $ comprises 
the user feature $\mathbf{x}_i$, 
the statistical feature $\mathbf{t}_i$, 
the historical behavior sequence $\mathbf{se}_i$,
and two observed responses: app usage time $y_i^{wt}$ and session frequency $y_i^{session}$. 
The historical behavior sequence is denoted as $\mathbf{se}_i = \{\mathbf{v}_{i,1}, \mathbf{v}_{i,2}, \dots, \mathbf{v}_{i,L}\}$, where $\mathbf{v}_{i,t}$ represents the behavior feature vector at the $t$-th historical step and $L$ is the sequence length.
The objective of L-RTG modeling is to estimate the user's expected total
consumption duration for the current day based on their historical behavior,
thereby providing accurate global guidance for GLAN inference.

\vspace*{-0.2cm}
\subsection{The Hierarchical Reward Modeling}
In the online video recommendation pipeline, we design a real-time data streaming for session-level evaluation.
Each instance is represented as a tuple $ (\mathbf{x}_i,\mathbf{c}_i,\mathbf{v}_i, \mathbf{k}_i,\mathbf{Y}_i)$, where $\mathbf{x}_i$ denotes the user features, 
$\mathbf{c}_i$ denotes the contextual features, 
$\mathbf{v}_i$ represents the corresponding session features, 
$\mathbf{k}_i$ indicates the landing page assigned by the platform 
and $\mathbf{Y}_i$ is the observed user response.
Specifically, $\mathbf{Y}_i = \{T_i^1, T_i^2, T_i^3, y_i^{\text{sw}}\}$, where $T_i^j$ denotes the consumption duration on the $j$-th page type.
The binary signal $y_i^{sw} \in \{0, 1\}$ indicates whether the user quickly abandons (drop off) the assigned landing page.
Formally, if the consumption duration of landing page within the session is less
than a threshold $\tau$ , the user is considered dissatisfied
with the assigned landing page and $y_i^{\text{sw}}$ is labeled as 1; otherwise, it is labeled as 0.
The objective of hierarchical reward modeling is to leverage these features
to accurately evaluate session-level value, thereby providing precise local
supervision signals for GLAN.
\section{Method}

\begin{figure*}
    \centering
    \includegraphics[width=0.9\linewidth]{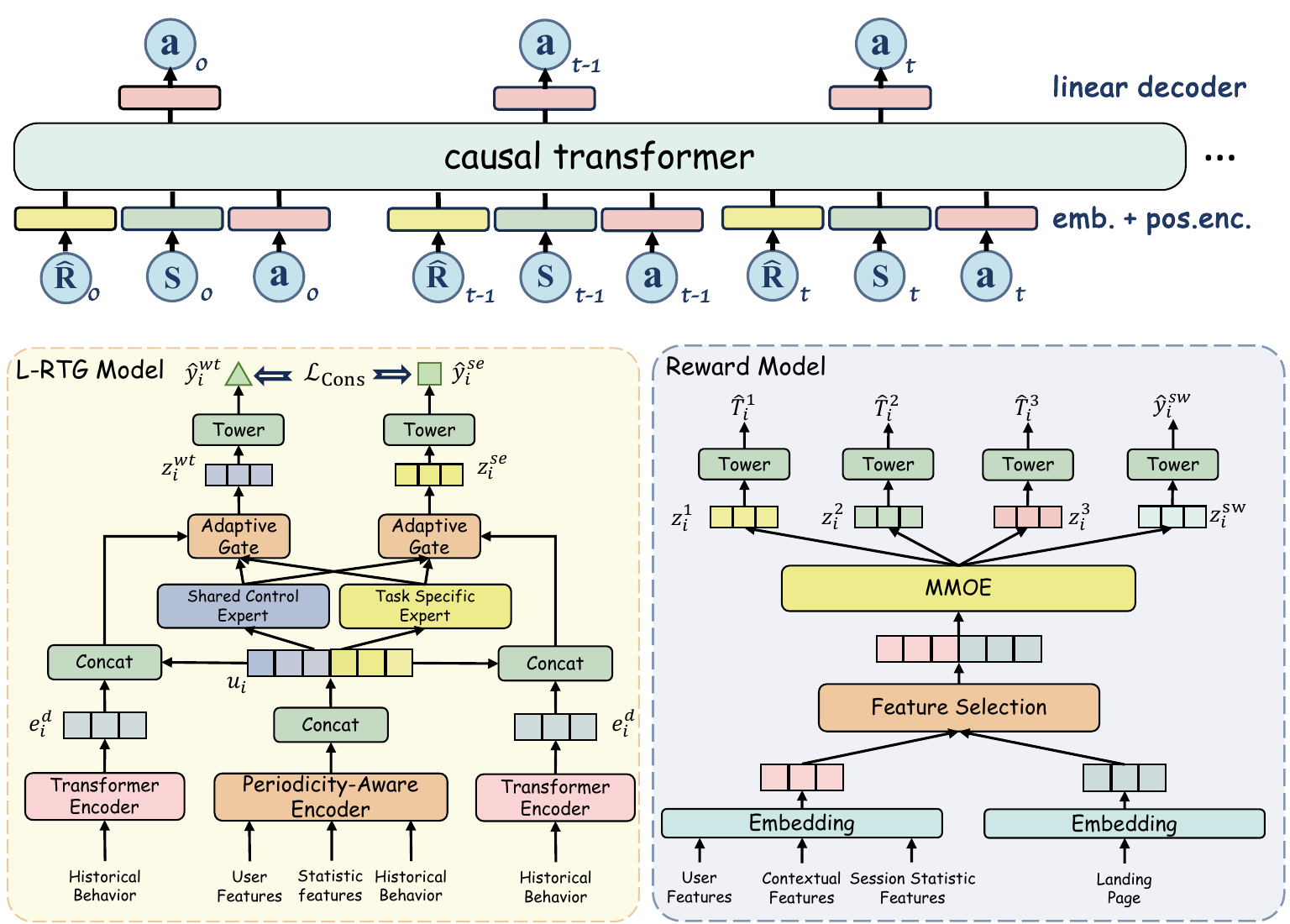}
    \vspace{-0.2cm}
    \caption{The architecture of our proposed method.}
    \label{fig:model}
    \vspace{-0.3cm}
\end{figure*}

\subsection{L-RTG}
\subsubsection{\textbf{Periodicity-Aware Feature Representation:}}
We first employ trainable embedding layers to obtain the representations of user features $\mathbf{x}_i$ and the statistical feature $\mathbf{t}_i$, denoted as $\mathbf{e}_{xi}$ and $\mathbf{e}_{ti}$, respectively. 
These representations are then concatenated to form a unified static context embedding.
The context embedding is formulated as:
\begin{equation}
    \mathbf{e}_i^{context} = \text{Concat}(\mathbf{e}_{xi}, \mathbf{e}_{ti}).
\end{equation}
For the historical behavior sequence $\mathbf{se}_i$, we apply a transformation layer to derive the corresponding latent representations $\mathbf{H}_i = [\mathbf{h}_{i,1}, \dots, \mathbf{h}_{i,L}] \in \mathbb{R}^{L \times d}$,
where $L$ denotes the sequence length and $d$ is the embedding dimension.

To aggregate the sequential information, we compute the attention~\cite{Yu_2020, vaswani2023attentionneed}, scores between the target context and the historical sequence. Unlike standard dot-product attention, we introduce a \textit{modulo-based periodic bias} to explicitly capture weekly periodicity inherent in user behavior patterns. The attention scores are computed as:
\begin{equation}
    \mathbf{w}_i = \frac{\mathbf{q}_i \mathbf{K}_i^\top}{\sqrt{d_k}} + \mathbf{b}_{period},
\end{equation}
where $\mathbf{q}_i = \mathbf{e}_i^{context} \mathbf{W}_Q \in \mathbb{R}^{d_k}$ denotes the query vector derived from the context embedding and $\mathbf{K}_i = \mathbf{H}_i \mathbf{W}_K \in \mathbb{R}^{L \times d_k}$ represents the key matrix of the historical sequence. Here, $\mathbf{W}_Q$ and $\mathbf{W}_K$ are learnable projection parameters. $\mathbf{b}_{\text{period}} \in \mathbb{R}^{L}$ represents the learnable periodicity bias:
\begin{equation}
\mathbf{b}_{\text{period}}[t] 
= \mathbf{B}_{\text{week}}\!\left[\Delta t \bmod 7\right] 
+ \mathbf{B}_{\text{dist}}\!\left[\text{bucket}(\Delta t)\right],
\end{equation}
where $\Delta t$ denotes the temporal distance between the current context and the $t$-th historical step. 
$\mathbf{B}_{\text{week}} \in \mathbb{R}^{7}$ and $\mathbf{B}_{\text{dist}} \in \mathbb{R}^{B}$ are learnable lookup tables weekly periodic effects and temporal decay, respectively.
The function $\text{bucket}(\cdot)$ maps $\Delta t$ into $B$ non-uniform bins with finer granularity for recent days and coarser granularity for distant ones.

Finally, we obtain the periodicity-aware sequence representation $\mathbf{e}_i^{p}$ via the weighted aggregation:
\begin{equation}
    \mathbf{e}_i^{p} = \text{softmax}(\mathbf{w}_i) \cdot \mathbf{V}_i,
\end{equation}
where $\mathbf{V}_i = \mathbf{H}_i \mathbf{W}_V \in \mathbb{R}^{L \times d_v}$ represents the value matrix, and $\mathbf{W}_V$ is the corresponding projection matrix.
Then we define the user representation as 
\begin{equation}
\setlength{\abovedisplayskip}{0.5pt}
\setlength{\belowdisplayskip}{0.5pt}
    \mathbf{u}_i = \text{Concat}(\mathbf{e}_i^{context},\mathbf{e}_i^{p})
\end{equation}

\subsubsection{\textbf{Sequential Dynamics Modeling}}
While the periodicity branch captures cyclical patterns, user behaviors also exhibit evolving trends and stochastic fluctuations caused by dynamic strategies in platform. 
To capture the sequential dynamics, we employ a standard Transformer Encoder to model the historical sequence dependencies:
\begin{equation}
\setlength{\abovedisplayskip}{0.5pt}
\setlength{\belowdisplayskip}{0.5pt}
\mathbf{H}_i^{d} = \text{Transformer\_Encoder}(\mathbf{H}_i + \mathbf{P}),
\end{equation}
where $\mathbf{P}$ denotes the absolute positional embeddings. We then apply an attention-pooling layer to aggregate the hidden states into a compact trend representation $\mathbf{e}_i^{d}$, which encapsulates the evolution of user interests and response variance:
\begin{equation}
\setlength{\abovedisplayskip}{0.5pt}
\setlength{\belowdisplayskip}{0.5pt}
\mathbf{e}_i^{d} = \text{AttentionPooling}(\mathbf{H}_i^{d}).
\end{equation}

\subsubsection{\textbf{Adaptive Gating Mechanism}}
Through the preceding modules, we obtain two complementary representations:
(1) The periodicity-aware user representation $\mathbf{u}_i$, , which encodes static user profiles and stable weekly cyclic patterns;
and (2) The sequential dynamics representation $\mathbf{e}_i^{d}$, modeling the evolving dynamics and stochastic fluctuations in user interests.

To enhance the model's representational capacity, we  first use operations (e.g. MLP) to generate $M$ expert representations $\{\mathbf{d_{im}}\}_{m=1}^M$ from $\mathbf{u}_i$.
To effectively aggregate the expert information, we introduce a task-specific adaptive gating mechanism.
Specifically, for each task $k \in \{wt, se\}$ (representing App Usage Time and Session, respectively), a dedicated gating network 
leverages both the sequential dynamics and periodicity-aware representations to compute task-specific aggregation weights:
\begin{equation}
\begin{aligned}
\mathbf{g}_i^k &= \text{Softmax}\left( \mathbf{W}_{g}^k (\mathbf{e}_i^{d} \mathbin{\|} \mathbf{u}_i ) + \mathbf{b}_{g}^k \right), \\
\mathbf{z}_i^k &= \sum_{m=1}^{M} g_{i}^k(m) \cdot \mathbf{d}_{im},
\end{aligned}
\label{eq:mmoe_fusion}
\end{equation}
where $W_g^k \in \mathbb{R}^{M \times d}$ and $b_g^k \in \mathbb{R}^{M \times 1}$ are transformation matrix and bias matrix, and $g_i^k \in \mathbb{R}^{M \times 1}$ represents the gating weights with $\mathbf{g}_i^k(m)$ denoting its $m$-th.element.
$z_i^k$ is the output latent representation of corresponding gate. 

The final score is then output through a specific tower, using this accumulated information as input, which can be formulated as:
\begin{equation}
\begin{aligned}
\hat{y}_i^{wt} &= \text{softplus}(h^{wt}(\mathbf{z}_i^{wt})), \\
\hat{y}_i^{se} &= \text{softplus}(h^{se}(\mathbf{z}_i^{se})),
\end{aligned}
\label{eq:dual_heads}
\end{equation}
where $h^k(\cdot)$ are the tower functions and the $\text{softplus}(\cdot)$ activation is applied to ensure non-negative outputs.

\subsubsection{\textbf{Optimization}}
Motivated by the inherent synergy between app usage time and session frequency in PLPM (as shown in Figure \ref{fig:DA}(a)), we formulate the training of L-RTG as a constrained optimization problem.
The primary objective is to prioritize the accuracy of app usage time prediction while constraining (i) the session-frequency prediction quality and (ii) the structural consistency between the two predictions:
\begin{equation}
\setlength{\abovedisplayskip}{0.5pt}
\setlength{\belowdisplayskip}{0.5pt}
\min_{\Theta}\ \bar{\mathcal{L}}_{wt}(\Theta)
\quad \text{s.t.}\quad
\bar{\mathcal{L}}_{se}(\Theta)\le \epsilon_{se},\ 
\bar{\mathcal{L}}_{rel}(\Theta)\le \epsilon_{rel},
\end{equation}
where $\bar{\mathcal{L}}_{k}(\Theta)=\frac{1}{|\mathcal{B}|}\sum_{i\in\mathcal{B}}\mathcal{L}_{k,i}$ denotes the mini-batch average loss over a batch $\mathcal{B}$.

We now specify each loss term. Given the long-tailed nature of app usage time distributions, we adopt the Huber loss to mitigate the influence of extreme values:
\begin{equation}
\mathcal{L}_{wt}(y_i^{wt}, \hat{y}_i^{wt}) = 
\begin{cases}
\frac{1}{2}(y_i^{wt} - \hat{y}_i^{wt})^2, & \text{if } |y_i^{wt} - \hat{y}_i^{wt}| \le \delta \\
\delta \left( |y_i^{wt} - \hat{y}_i^{wt}| - \frac{1}{2}\delta \right), & \text{otherwise}
\end{cases}
\end{equation}
where $\delta$ is a threshold hyperparameter. For session frequency, we use the standard Mean Squared Error (MSE) loss:
\begin{equation}
\setlength{\abovedisplayskip}{0.5pt}
\setlength{\belowdisplayskip}{0.5pt}
\mathcal{L}_{se}(y_i^{se}, \hat{y}_i^{se}) = (y_i^{se} - \hat{y}_i^{se})^2.
\end{equation}
The structural-consistency loss enforces the empirical relationship between the two targets observed in online data (Figure~\ref{fig:DA}(a)):
\begin{equation}
\setlength{\abovedisplayskip}{0.5pt}
\setlength{\belowdisplayskip}{0.5pt}
\mathcal{L}_{rel,i}=\left(\hat{y}_i^{wt}-a\ln(1+\hat{y}_i^{se})-b\right)^2.
\end{equation}
Here $a$ and $b$ are obtained from online data statistics.

We solve the constrained problem via primal-dual optimization of the Lagrangian:
\begin{equation}
\setlength{\abovedisplayskip}{0.5pt}
\setlength{\belowdisplayskip}{0.5pt}
\min_{\Theta}\max_{\lambda_{se}\ge 0,\lambda_{rel}\ge 0}\ 
\bar{\mathcal{L}}_{wt}(\Theta)
+\lambda_{se}\left(\bar{\mathcal{L}}_{se}(\Theta)-\epsilon_{se}\right)
+\lambda_{rel}\left(\bar{\mathcal{L}}_{rel}(\Theta)-\epsilon_{rel}\right).
\end{equation}
At each iteration, the model parameters $\Theta$ are updated by gradient descent, while the dual variables are updated by projected gradient ascent:
\begin{equation}
\setlength{\abovedisplayskip}{0.5pt}
\setlength{\belowdisplayskip}{0.5pt}
\lambda_{se}\leftarrow \left[\lambda_{se}+\eta_{\lambda}\left(\bar{\mathcal{L}}_{se}-\epsilon_{se}\right)\right]_+,\quad
\lambda_{rel}\leftarrow \left[\lambda_{rel}+\eta_{\lambda}\left(\bar{\mathcal{L}}_{rel}-\epsilon_{rel}\right)\right]_+,
\end{equation}
where $[x]_+=\max(0,x)$. To reduce the variance of mini-batch estimates and stabilize the dual updates, we maintain an exponential moving average (EMA) of the constraint losses:
\begin{equation}
\setlength{\abovedisplayskip}{0.5pt}
\setlength{\belowdisplayskip}{0.5pt}
\tilde{\mathcal{L}}_{k}\leftarrow \rho\,\tilde{\mathcal{L}}_{k}+(1-\rho)\,\bar{\mathcal{L}}_{k}, \quad k\in\{se,rel\},
\end{equation}
and substitute $\tilde{\mathcal{L}}_{k}$ for $\bar{\mathcal{L}}_{k}$ in the dual updates. In practice, we also clip the dual variables to $\lambda_k \le \lambda_{\max}$ to prevent unbounded penalty growth.

\subsubsection{\textbf{Inference}}
During inference, the L-RTG module predicts the expected app usage time $\hat{y}_i^{wt}$, which is subsequently adopted by GLAN as the initial return-to-go $\hat{R}_1$, providing a global guidance signal for the day-level policy.

\subsection{Hierarchical Reward Model}
\subsubsection{\textbf{Feature Selection and Fusion}}
Given an input instance  $ (\mathbf{x}_i,\mathbf{c}_i,\mathbf{v}_i, \mathbf{k}_i,\mathbf{Y}_i)$, we first use trainable embedding layers to obtain the representations of all of them, denoted as $\mathbf{e}_{xi}$, $\mathbf{e}_{ci}$, $\mathbf{e}_{vi}$ and $\mathbf{e}_{ki}$, respectively.
We then apply a target attention mechanism for page-specific feature selection, extracting relevant information from $\mathbf{e}_{xi} \| \mathbf{e}_{ci} \| \mathbf{e}_{vi}$ guided by the landing page embedding $\mathbf{e}_{ki}$:
\begin{equation}
\setlength{\abovedisplayskip}{0.5pt}
\setlength{\belowdisplayskip}{0.5pt}
\mathbf{e}_{ui} = \text{FS}((\mathbf{e}_{xi} || \mathbf{e}_{ci} || \mathbf{e}_{vi}), \mathbf{e}_{ki}),
\end{equation}
where $\text{FS}(\cdot)$ denotes the page-specific feature selection module (detailed in Appendix \ref{sec:feature_selection}). 
These embeddings are then concatenated to form $\mathbf{f}_i$, which serves as the input feature vector for the subsequent stage:
\begin{equation}
\setlength{\abovedisplayskip}{0.5pt}
\setlength{\belowdisplayskip}{0.5pt}
    \mathbf{f}_i = \text{Concat}(\mathbf{e}_{ui}, \mathbf{e}_{ki})
\end{equation}

\subsubsection{\textbf{Multi-Task Prediction}}
To effectively model the correlations among duration distributions of different page types while isolating the landing page's specific risk (i.e., immediate Landing-page switching), we adopt the Multi-gate Mixture-of-Experts (MMoE) as the shared backbone:
\begin{equation}
\setlength{\abovedisplayskip}{1pt}
\setlength{\belowdisplayskip}{1pt}
\mathbf{z}^{1}, \mathbf{z}^{2}, \mathbf{z}^{3}, \mathbf{z}^{\text{sw}} = \text{MMoE}(\mathbf{f_i}).
\end{equation}
Subsequently, four independent towers process these representations to generate the final predictions:
\begin{equation}
\begin{aligned}
    \hat{T}_{i}^k &= \text{softplus}(h^k(\mathbf{z}_i^{k})), \quad k \in \{1,2,3\} \\
    \hat{y}_{i}^{\text{sw}} &= \sigma(h^{\text{sw}}(\mathbf{z}_i^{\text{sw}})),
\end{aligned}
\end{equation}
where $h^k(\cdot)$ and $h^{\text{sw}}(\cdot)$ denote the task-specific tower networks (e.g., MLPs), and $\sigma(\cdot)$ is the sigmoid function.

\subsubsection{\textbf{Optimization}}
Through the preceding steps, we obtain the estimated outputs $[\hat{T}_{i}^1, \hat{T}_{i}^2, \hat{T}_{i}^3, \hat{y}_{i}^{sw}]$.
For training
, we design a hybrid loss comprising a \textit{fine-grained regression loss} and a \textit{page drop-off risk penalty}.

\textbf{Fine-grained Regression Loss.}
To capture the fine-grained signal within each session, we supervise duration predictions across all page types using the Huber Loss:
\begin{equation}
\mathcal{L}_{\text{reg}} = \sum_{j=1}^{3} \text{Huber}(\hat{T}_i^{j}, T_i^{j}).
\end{equation}

\textbf{page drop-off risk penalty.}
The switching label $y_{i}^{\text{sw}}$ is highly imbalanced in the online data stream.
Moreover, false negatives (predicting retention when the user actually switches) are particularly detrimental, as they encourage the dispatching of mismatched pages.
We therefore adopt a focal binary cross-entropy loss to emphasize hard samples and penalize overconfident predictions:
\begin{equation}
\mathcal{L}_{\text{sw}}
=
-\frac{1}{|\mathcal{B}|}\sum_{i\in\mathcal{B}}
\Big[
y_i^{\text{sw}}(1-\hat{y}_i^{\text{sw}})^{\gamma}\log(\hat{y}_i^{\text{sw}})
+
(1-y_i^{\text{sw}})(\hat{y}_i^{\text{sw}})^{\gamma}\log(1-\hat{y}_i^{\text{sw}})
\Big],
\end{equation}
where $\gamma$ is hyperparameter and $\mathcal{B}$ denotes the training batch.

The overall loss for hierarchical reward modeling (HRM) is:
\begin{equation}
\mathcal{L}_{\text{total}} = \mathcal{L}_{\text{reg}} + \lambda \cdot \mathcal{L}_{\text{sw}}.
\end{equation}
where $\lambda$ balances the regression and drop-off objectives.

\subsubsection{\textbf{Inference}}
During inference, the session-level reward is computed as the expected total duration calibrated by switch risk:
\begin{equation}
\setlength{\abovedisplayskip}{0.5pt}
\setlength{\belowdisplayskip}{0.5pt}
\hat{r}_i = (1 - \hat{y}_{i}^{\text{sw}}) \cdot \sum_{j=1}^{3} \hat{T}_i^{j}.
\end{equation}
This formulation discounts sessions with high drop-off risk, providing precise local supervision for GLAN.

\section{Experiment}

\subsection{Dataset}

\subsubsection{\textbf{Daily-level Data}}
Here, we evaluate our L-RTG using an industrial-scale dataset collected from the Kuaishou short-video platform. The dataset consists of two months of anonymized user interaction logs, tailored for the task of daily-level Return-to-Go (RTG) prediction. We will publicly release this large-scale real-world benchmark to promote further research in this area upon paper acceptance.
Specifically, each instance includes three categories of features: 
(1) user profile attributes such as age, region, and activity level; 
(2) statistical consumption features, including general consumption statistics (e.g., app usage duration, video view counts) over the past 7 and
30 days and page-specific engagement metrics(e.g., page stay duration, page video playtime, and page livestream playtime) over the past 7 and 14 days; 
and 
(3) Dynamic behavior sequence over the past 30 days, including daily session frequency and daily app usage time.
As for the ground-truth labels, we collect both the total app usage time and the session frequency of the following day.

\subsubsection{\textbf{Session-level Data}}
To facilitate precise session value evaluation and reward model training,
we construct a real-time stream dataset derived from online user interaction logs.
Each instance encompasses a complete session-level page sequence $\xi_t$, annotated with fine-grained page-specific consumption features (e.g., watch time and video view counts).
To enable real-time user state modeling, we incorporate two complementary categories of features during training: 
(1)real-time contextual features, including current-day page-level consumption statistics, the number of live broadcasts by followed creators at the time of session entry, and the count of prior app entries on the same day; 
and (2) historical statistical features, comprising aggregated page-level consumption behaviors over the past $n$ days.

\begin{figure}[!t]
	\centering
	\setlength{\belowcaptionskip}{-0.0cm}
	\setlength{\abovecaptionskip}{-0.0cm}
	\includegraphics[width=\linewidth]{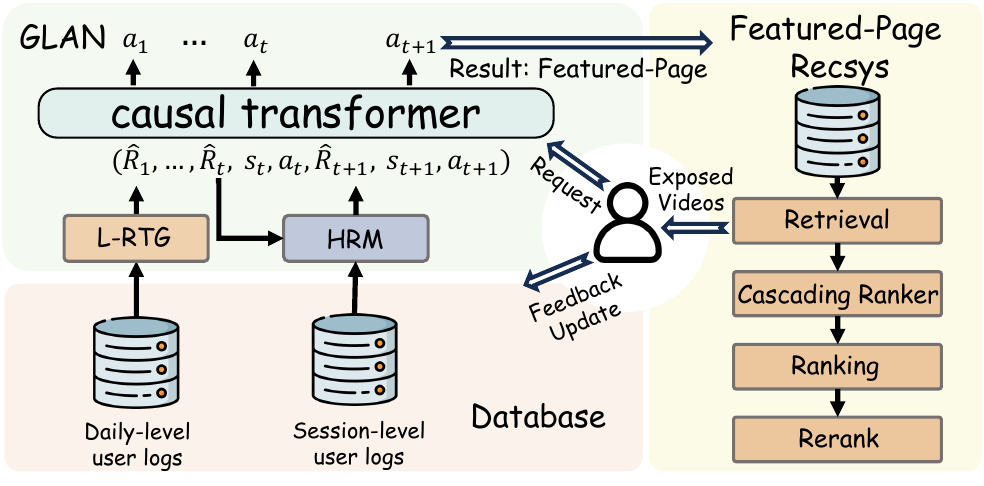}
    \vspace{-0.4cm}
\caption{Illustration of the online pipeline.}
	\label{fig:online_pipeline}
    \vspace{-0.7cm}
\end{figure}

\subsection{Online A/B Test}
To further evaluate the online performance of our proposed GLAN in a real-world production environment, we rebuilt the original personalized landing page service and directly deployed GLAN on the Kuaishou platform, which serves hundreds of millions of users. We then conducted an online experiment on approximately 5\% of the traffic for about 56 days, covering both the AA and AB testing phases.

\subsection{\textbf{System Description}}
As shown in Figure \ref{fig:online_pipeline}, we present the online serving pipeline of the Kuaishou platform.
When a user enters the app, a request is dispatched to the server, which first invokes the personalized landing page service. 
in the pipeline of the Kuaishou platform. 
This service processes the user's historical interaction records, intra-day session trajectory, and other contextual information to determine the optimal landing page. 
Once the page indicator is returned, the initial request completes, and subsequent requests are triggered to invoke the corresponding downstream services 
Taking the featured-page recommendation service as an example, this request will send the corresponding user attributes and context features to the featured-page recommendation online service. 
Then trigger the online pipeline to filter and select videos from the candidate pool in a cascading form. 
Finally, after guiding the user to the corresponding page, the selected videos will be displayed to the user.  
Last but not least, when the user kills an app process in the background and re-enters the app, a new request is initiated to invoke the personalized landing page service once again, thereby beginning a new session.
The current online baseline is instantiated by our previously deployed framework, \textbf{KLAN}, which employs a Conservative Q-Learning (CQL) model updated on an hourly basis. 
For each user request, the CQL-based model estimates the Q-values of candidate pages and greedily selects the page with the maximum expected reward.
The inherent bootstrapping instability and the absence of global trajectory modeling have made the long-term value estimation increasingly unreliable, posing significant challenges for precise credit assignment and daily-level optimization.

\subsubsection{\textbf{Scheme Overview.}}
Our proposed framework is built upon the \textbf{Decision Transformer (DT)}, augmented with two key modules: the \textbf{L-RTG Module} for initializing the target return condition, and the \textbf{HRM} for session-level value evaluation.
For each user within a day, the inference proceeds as follows:
\noindent\textbf{(1) Initial RTG Setting.}
At the beginning of each day, the L-RTG module is invoked \textit{once} to initialize the target return. Specifically, it analyzes the user's inter-day profile to predict a personalized target Return-to-Go, denoted as $\hat{R}_1$. This value serves as the initial conditioning signal for GLAN.
\noindent\textbf{(2) Autoregressive Action Generation.}
Upon each app entry, GLAN generates a landing page action. 
At step $t$, it takes the current state $s_t$, the remaining target return $\hat{R}_t$ and the historical trajectory $\tau_{<t}$ as input.
Leveraging its causal self-attention mechanism, GLAN autoregressively generates the optimal landing page action:
\begin{equation}
    a_t \sim \pi_{\theta}(\cdot \mid s_t, \hat{R}_t, \tau_{<t}),
\end{equation}
where $\tau_{<t}$ denotes the sequence of past states, actions, and RTG within the current day.
\noindent\textbf{(3) Session-level Reward Evaluation and RTG Update.}
Upon session completion, 
HRM module evaluates the user's realized behavior and computes a scalar reward $\hat{r}_t$.
The remaining target return is then updated accordingly:
\begin{equation}
    \hat{R}_{t+1} = \hat{R}_t - \hat{r}_t
\end{equation}
\noindent Steps (2) and (3) alternate iteratively throughout the day: each new app entry triggers action generation conditioned on the updated RTG, followed by reward evaluation upon session completion. This closed-loop process continues until no further sessions occur, enabling the Decision Transformer to dynamically adjust its policy across sessions to fulfill the daily engagement objective.

\begin{table}[t]
   \setlength{\abovecaptionskip}{0cm}
   \setlength{\belowcaptionskip}{-0.0cm}

\caption{The performance of Online A/B Testing at the platform. Boldface represents significance level $p$-value $<0.05$ of comparing GLAN with the online baseline. The values presented in the table represent the average observations of the last 7 days.}
    \centering
	\resizebox{0.88\linewidth}{!}{
    \begin{tabular}{c|c}
    \toprule
    {Dataset}&
    \multicolumn{1}{c}{Online Platform} \cr
    \cmidrule(lr){1-1}
    \cmidrule(lr){2-2}
    {Metrics} &\textbf{DT}\cr
    \midrule
    Daily Active Users &\textbf{+0.158\%}\cr
    Lifetime (LT) &\textbf{+0.108\%}\cr
    APP usage time &\textbf{+0.369\%}\cr
    Watch Time &\textbf{+0.394\%}\cr
    Video View &\textbf{+0.469\%}\cr
    Page Drop-off Ratio &\textbf{-15.832\%}\cr
    Overall Page Effective Entry Frequency &\textbf{+1.079\%}\cr
    Latency &+0.087\%\cr
    \bottomrule
    \end{tabular}}
    \label{online_ab111}
    \vspace{-6mm}
\end{table}

\begin{figure*}[t]
   \setlength{\abovecaptionskip}{0.2cm}
   \setlength{\belowcaptionskip}{-0.0cm}
    \centering     \vspace{-0.25cm}
    \includegraphics[width=1.00\linewidth]{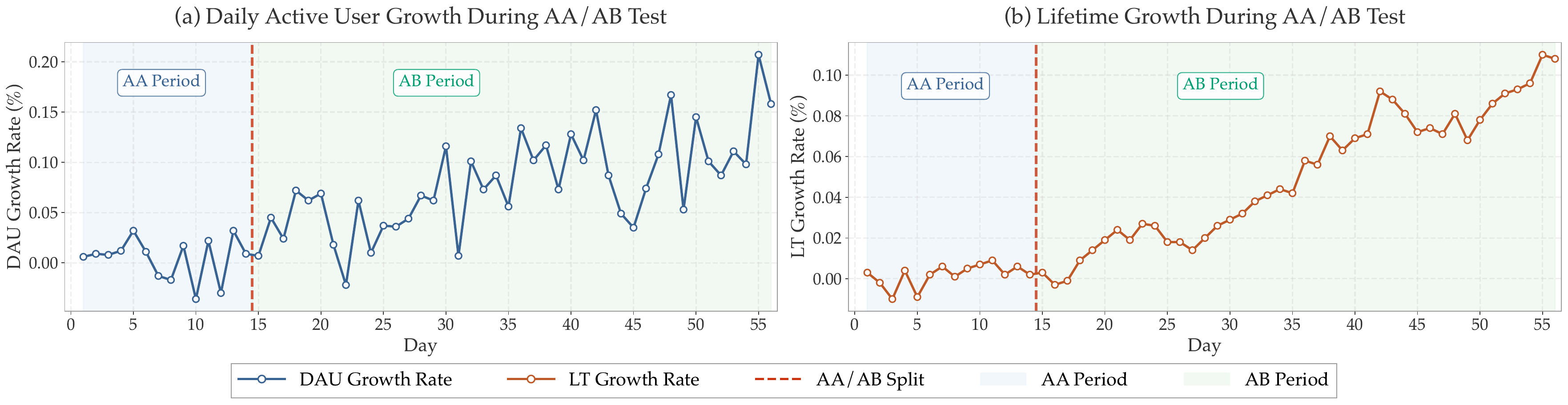}
    \vspace{-0.4cm}
    \caption{Online A/B test experimental results of DAU and LT.}
    \label{fig:online}
    \vspace{-0.25cm}
\end{figure*}

\subsection{Online Experimental Results}
\subsubsection{\textbf{Online A/B test results}}
We evaluate GLAN using common consumption metrics (APP usage time, watch time, and daily active users), commercial metrics (video view), PLPM-specific metrics (page drop-off ratio), platform constrain metrics (effective page entry frequency and latency), and retention metrics (LT) to comprehensively assess its online performance.
Among these metrics, \textbf{LT} serves as a critical indicator of user retention and experience quality \cite{meng2024coarsetofinedynamicupliftmodeling}, defined as:
\begin{equation}
\setlength{\abovedisplayskip}{1pt}
\setlength{\belowdisplayskip}{1pt}
    \text{LT} = \frac{\sum_{i = \max(T - 6, T_0)}^T \text{DAU}_i}{\text{Total number of active users from } T_0 \text{ to } T}
    \label{equ:LT7}
\end{equation}
Figure \ref{fig:online} illustrates the temporal evolution of DAU and LT growth rates throughout the 56-day experimental period, comprising a 14-day AA test followed by a 42-day AB test.
During the AB test phase, GLAN demonstrates consistent and significant improvements in both metrics. 
Table~\ref{online_ab111} reports statistically significant gains across all evaluation dimensions compared to the online baseline.
Notably, APP usage time, watch time and video view counts show substantial growth, while page drop-off ratio decreases significantly, indicating enhanced user satisfaction with assigned landing pages. 
Regarding platform constraint metrics, the effective page entry frequency metrics for all three pages show positive improvements.
Latency results indicate our model maintains real-time efficiency with negligible overhead.
Currently, GLAN is fully deployed on Kuaishou, serving hundreds of millions of daily active users every day.

\begin{table}[t]
   \setlength{\abovecaptionskip}{0.1cm}
   \setlength{\belowcaptionskip}{-0.0cm}
\caption{The ablation result of Online A/B Testing at the platform. Boldface represents significance level $p$-value $<0.05$ of comparing GLAN variants with GLAN.}
    \centering
	\resizebox{0.85\linewidth}{!}{
    \begin{tabular}{c|ccc}
    \toprule
    {Scenario}&
    \multicolumn{3}{c}{Online Platform} \cr
    \cmidrule(lr){1-1}
    \cmidrule(lr){2-4}
    {Metrics} &APP usage time &Watch Time &Video View\cr
    \midrule
    GLAN & \textbf{+0.369\%}&\textbf{+0.394\%} &\textbf{+0.469\%} \cr
    w/o L-RTG &\textbf{+0.137\%}&\textbf{+0.175\%}&\textbf{+0.158\%}\cr
    w/o HRM &\textbf{+0.277\%}&\textbf{+0.278\%}&\textbf{+0.404\%}\cr
    \bottomrule
    \end{tabular}}
    \label{online_ab222}
    \vspace{-14pt}
\end{table}

\subsubsection{\textbf{Ablation Test of GLAN}}
We conduct online ablation studies to evaluate the contribution of each key component in GLAN.
We compare GLAN against two degraded variants:
(1) \textbf{w/o L-RTG}, which removes the L-RTG module and initializes the target RTG using the user's historical average engagement over the past 7 days;
(2) \textbf{w/o HRM}, which replaces the hierarchical reward model with a naive reward defined as the total session duration.
As shown in Table~\ref{online_ab222}, all DT-based variants consistently outperform the online baseline (KLAN), demonstrating the effectiveness of modeling intra-day session trajectories from a global perspective through the sequence modeling paradigm.
Regarding individual components, \textbf{w/o L-RTG} incurs the largest performance drop, indicating that static heuristics fail to capture inter-day engagement volatility and result in inaccurate global guidance that hinders optimal policy generation.
Removing HRM also leads to notable degradation across all metrics, confirms that coarse-grained session-level rewards are insufficient for precise credit assignment.
Without hierarchical decomposition, the model struggles to distinguish between a genuinely effective landing page assignment and a suboptimal assignment compensated by subsequent user actions.
The integration of global guidance (via L-RTG) and local supervision (via HRM) enables synergistic performance beyond their individual contributions.

\subsubsection{\textbf{In-depth Analysis}}
To further investigate the source of performance gains, we analyze the online results of landing-page assignment distribution and the corresponding user engagement across page types (Table~\ref{indepth-analysis}).
Historically, the Featured Page dominates the assignment of landing page due to its generally high consumption duration, leading to a "winner-takes-all" phenomenon in online baseline.
GLAN significantly alleviates this imbalance: the Featured Page assignment ratio decreases, while the ratios for the Explore and Following pages substantially increase.
This indicates that GLAN, leveraging the global receptive field of the Decision Transformer, effectively breaks the greedy feedback loop and uncovers diverse user intents that were previously underserved.
Crucially, the effective page entry frequency metrics for all three pages show positive improvements, confirming that the reassigned traffic genuinely matches user preferences rather than merely diversifying assignments.

\begin{table}[!t]
   \setlength{\abovecaptionskip}{0.1cm}
   \setlength{\belowcaptionskip}{-0.0cm}

\caption{
The page-specific results of Online A/B Testing at the platform. Boldface represents significance level $p$-value $<0.05$ of comparing GLAN with the online baseline.}
    \centering
	\resizebox{0.88\linewidth}{!}{
    \begin{tabular}{c|c}
    \toprule
    {Dataset}&
    \multicolumn{1}{c}{Online Platform} \cr
    \cmidrule(lr){1-1}
    \cmidrule(lr){2-2}
    {Metrics} &\textbf{DT}\cr
    \midrule
    Featured Page Assign Ratio &\textbf{-2.481\%}\cr
    Explore Page Assign Ratio &\textbf{+5.264\%}\cr
    Following Page Assign Ratio &\textbf{+8.461\%}\cr
    Featured Page Effective Entry Frequency &\textbf{+0.825\%}\cr
    Explore Page Effective Entry Frequency &\textbf{+3.717\%}\cr
    Following Page Effective Entry Frequency &\textbf{+2.489\%}\cr
    \bottomrule
    \end{tabular}}
    \label{indepth-analysis}
\end{table}

\section{Conclusion and Future Work}
In this paper, we formally identify the fundamental limitations inherent in the existing CQL-based KLAN framework for PLPM.
To address it, we propose GLAN, an enhanced DT
framework designed to provide personalized landing pages under
the formulation of PLPM. 
Extensive online experiments on Kuaishou demonstrate GLAN's effectiveness and scalability. 
Moreover, our proposed GLAN is now deployed online with full users at Kuaishou platform, serving for hundreds of millions of users every day.
\normalem
\bibliographystyle{ACM-Reference-Format}
\balance
\bibliography{reference}
\nocite{*}

\newpage

\appendix

\section{APPENDIX}

\begin{figure}[H]
	\centering
	\subfigure[Featured Page.]{\includegraphics[height=.6\linewidth]{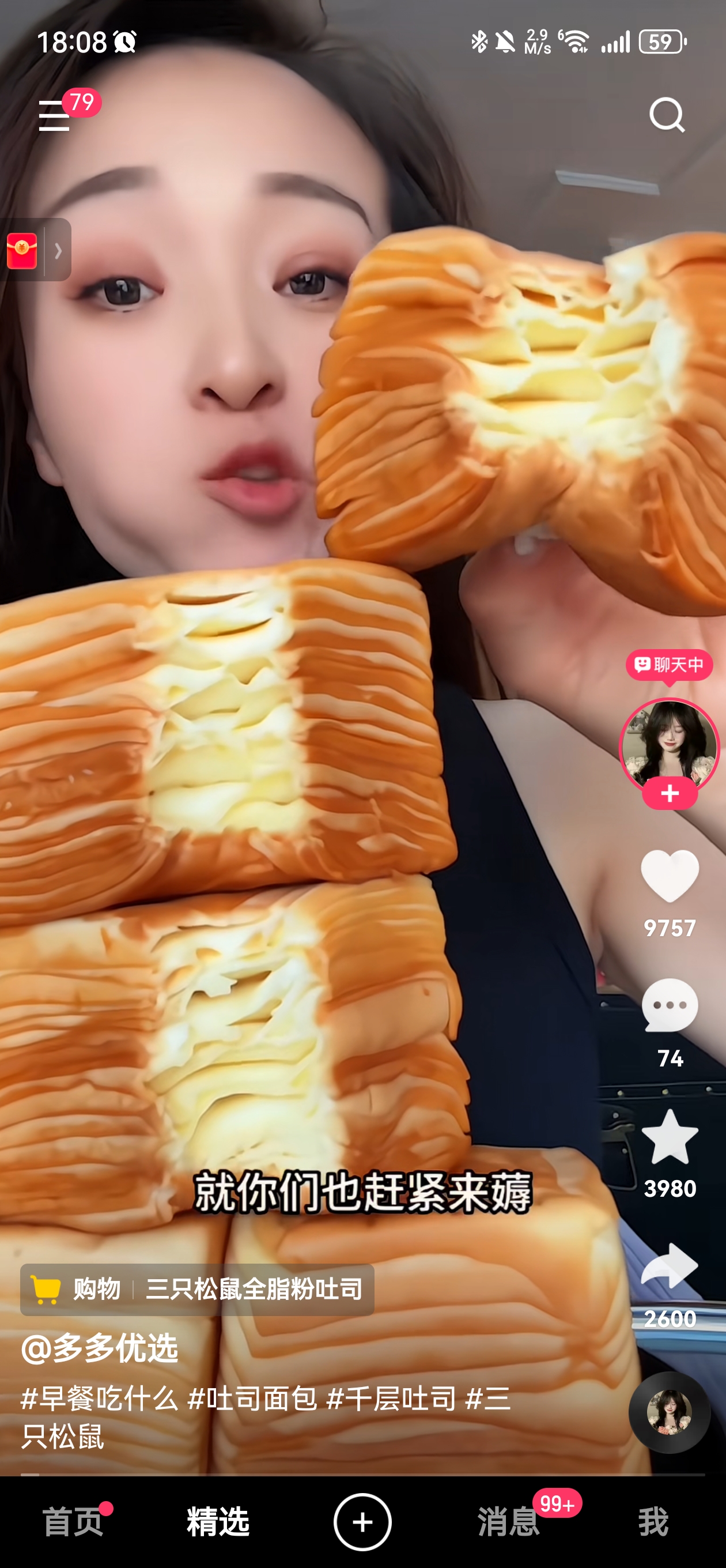}}\hspace{0.3cm} 
	\subfigure[Store Page.]{\includegraphics[height=.6\linewidth]{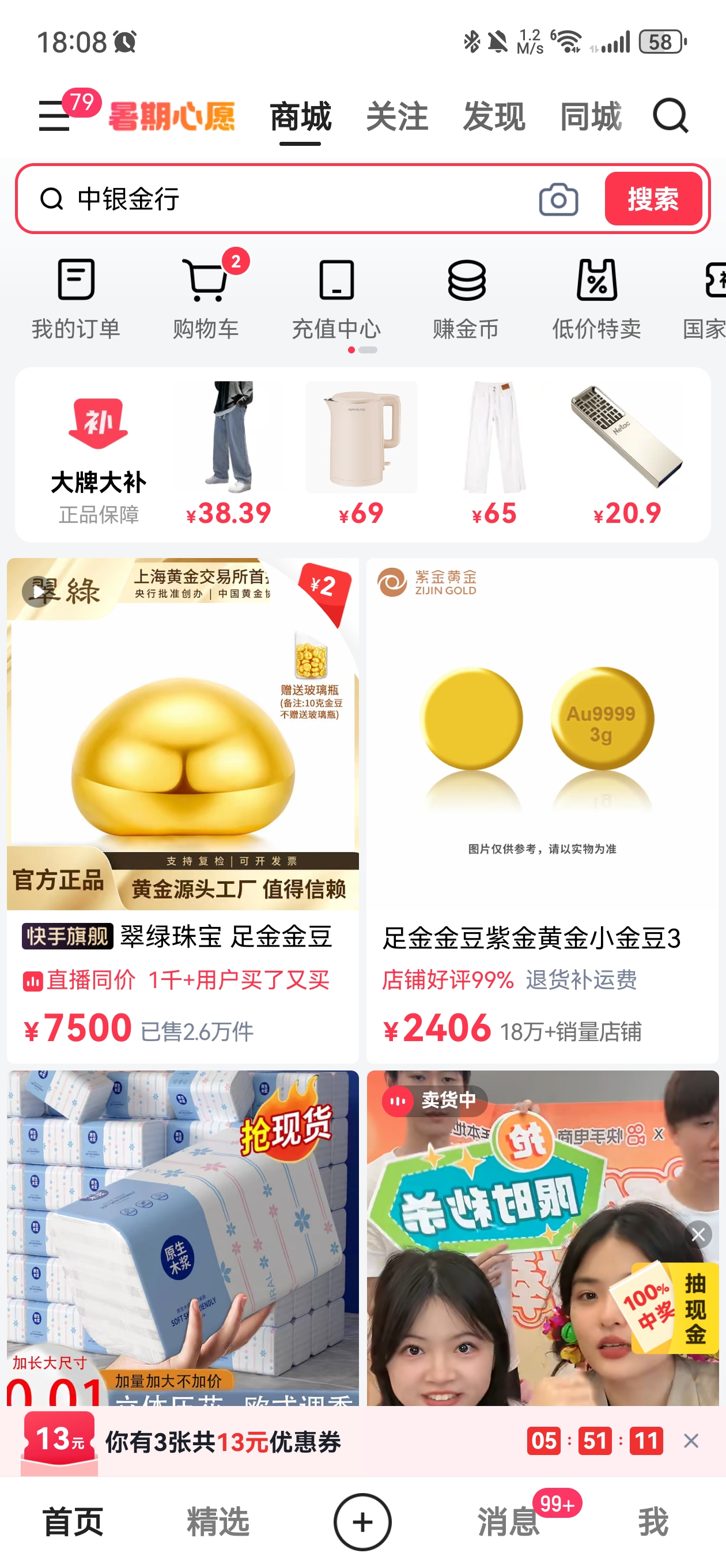}}\hspace{0.3cm} 
    \subfigure[Following Page.]{\includegraphics[height=.6\linewidth]{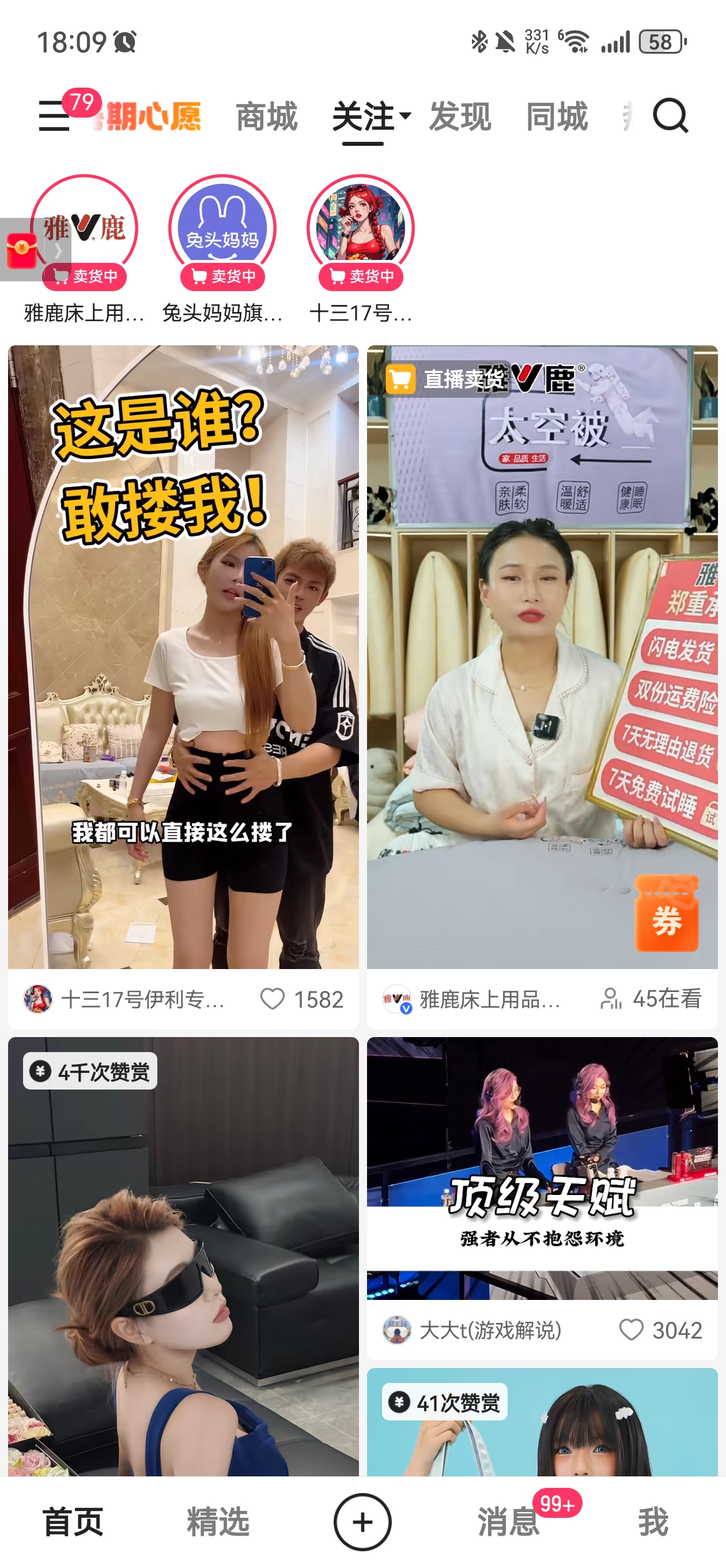}}
	\vspace{-0.4cm}
	\caption{Cases of different pages on Kuaishou APP.}
	\label{fig:platform_description}
	\vspace{-0.25cm}
\end{figure}

\subsection{Platform Description}
\label{sec:page_heterogeneity}
To satisfy different users interests, Kuaishou APP offers multiple functional and channel pages with distinct characteristics, leading to significant page heterogeneity.
This heterogeneity spans multi-dimensions (layout, content, and interaction patterns) and results in vastly different user response distributions across pages. 
Figure \ref{fig:platform_description} illustrates three representative page types that demonstrate this principle.

\textbf{Featured Page} employs a full-screen, single-column layout optimized for immersive content consumption. 
Users can scroll vertically to seamlessly transition between videos, while strategically placed interaction buttons (like, comment, follow) encourage real-time engagement. 
In contrast, \textbf{Store Page} adopts a double-column layout, similar to traditional e-commerce platforms, prioritizing commercial efficiency over entertainment. Clear product listings with prices enable users to quickly browse and compare multiple items. The interface emphasizes transactional actions—clicking products, viewing details, adding to cart—reflecting its commerce-oriented objectives.
\textbf{Following Page} strikes a balance between the two approaches, utilizing a double-column format that presents content from following creators. 
This layout optimizes for both content discovery and social connection, allowing users to efficiently browse updates and live streaming information from familiar creators while maintaining the personal touch through likes and comments. 
Such page diversity, while essential for user experience, introduces substantial complexity for recommendation systems. The divergent user behaviors and response patterns across these pages challenge traditional joint-learning approaches, necessitating more sophisticated modeling strategies to effectively capture page-specific preferences.

\subsection{Feature Embedding and Selection}
\label{sec:feature_selection}
Given an input instance $(\mathbf{x}_i, \mathbf{c}_i, \mathbf{v}_i, \mathbf{k}_i, \mathbf{Y}_i)$, we first transform all features into dense embeddings through trainable embedding layers:
\begin{equation}
\mathbf{e}_{xi} = \mathbf{E}_x^\top \mathbf{x}_i, \quad
\mathbf{e}_{ci} = \mathbf{E}_c^\top \mathbf{c}_i, \quad
\mathbf{e}_{vi} = \mathbf{E}_v^\top \mathbf{v}_i, \quad
\mathbf{e}_{ki} = \mathbf{E}_k^\top \mathbf{k}_i,
\end{equation}
where $\mathbf{E}_x \in \mathbb{R}^{f_x \times d}$, $\mathbf{E}_c \in \mathbb{R}^{f_c \times d}$, $\mathbf{E}_v \in \mathbb{R}^{f_v \times d}$, and $\mathbf{E}_k \in \mathbb{R}^{f_k \times d}$ are the embedding tables for user, context, session, and landing-page features, respectively, and $d$ is the embedding dimension.

To perform page-specific feature selection, we employ a target attention mechanism that extracts relevant information from the concatenated non-page embeddings $\mathbf{e}_{xi} \| \mathbf{e}_{ci} \| \mathbf{e}_{vi}$ guided by the landing-page embedding $\mathbf{e}_{ki}$. Let $F = f_x + f_c + f_v$ denote the total number of non-page feature fields and $\mathbf{e}_{oi} = [\mathbf{e}_{oi,1}, \dots, \mathbf{e}_{oi,F}]$ denote the concatenated field-level embeddings. The selection mechanism is defined as:
\begin{equation}
\begin{aligned}
\mathbf{w}_i &= \text{Softmax}\!\left(\mathbf{W}^k \cdot \mathbf{e}_{ki} + \mathbf{b}^k\right), \\
\mathbf{e}_{ui} &= \sum_{j=1}^{F} \mathbf{w}_i(j) \cdot \mathbf{e}_{oi,j},
\end{aligned}
\end{equation}
where $\mathbf{W}^k \in \mathbb{R}^{F \times d}$ and $\mathbf{b}^k \in \mathbb{R}^{F}$ are page-specific learnable parameters, and the superscript $k$ indexes the landing page indicated by $\mathbf{k}_i$. The attention weight $\mathbf{w}_i \in \mathbb{R}^{F}$ acts as a soft selector, with $\mathbf{w}_i(j)$ determining the relevance of the $j$-th feature field to the assigned landing page. The resulting $\mathbf{e}_{ui}$ is the page-specific user representation used in the main text.

It is worth noting that for landing pages with similar semantics, the transformation parameters $\mathbf{W}^k$ and $\mathbf{b}^k$ are shared across the corresponding page branches to improve parameter efficiency.


\subsection{Computational Complexity Discussion}
We analyze the computational complexity of GLAN's three core components during both training and inference.

\subsubsection{\textbf{L-RTG Module}}
The periodicity-aware attention over the historical behavior sequence of length $L$ costs $\mathcal{O}(Ld)$, where $d$ is the embedding dimension.
The Transformer Encoder for sequential dynamics modeling incurs $\mathcal{O}(L^2 d)$ due to self-attention.
The adaptive gating mechanism with $M$ experts requires $\mathcal{O}(Md)$.
Since $M \ll L$, the overall complexity of L-RTG is $\mathcal{O}(L^2 d)$.
Crucially, L-RTG is invoked only \textit{once per user per day}, making its amortized per-session cost negligible.

\subsubsection{\textbf{HRM Module}}
The feature selection module computes a single-query attention over $F$ feature fields, costing $\mathcal{O}(Fd)$.
The MMoE backbone with $M'$ experts and 4 task-specific towers requires $\mathcal{O}(M'd^2)$.
The total per-session complexity is $\mathcal{O}(Fd + M'd^2)$, which is independent of the trajectory length and thus scales efficiently with user activity.

\subsubsection{\textbf{Decision Transformer}}
At step $t$, the causal self-attention operates over the trajectory of length $t$, with each token being an (RTG, state, action) triplet.
The per-step complexity is $\mathcal{O}(3t \cdot d)$ for a single attention layer, yielding $\mathcal{O}(td)$.
Across all $N$ sessions within a day, the total cost is $\mathcal{O}(\sum_{t=1}^{N} td) = \mathcal{O}(N^2 d)$.
In practice, since the typical daily session count $N$ is moderate (usually $N < 100$) and $|\mathcal{A}| = 3$, this remains highly efficient.
Furthermore, KV-cache can be employed during autoregressive inference to reduce the incremental cost to $\mathcal{O}(Nd)$ per step.

\subsubsection{\textbf{Overall}}
The total daily inference cost per user is $\mathcal{O}(L^2 d + N \cdot (Fd + M'd^2) + N^2 d)$.
Given that $L$, $N$, $F$, and $M'$ are all bounded constants in the PLPM setting, the framework maintains real-time efficiency, as empirically validated by the negligible latency overhead ($+0.087\%$) observed in our online deployment.

\end{document}